\documentclass[twoside]{article}
\usepackage{PRIMEarxiv}

\usepackage{cite}
\usepackage{amsmath,amssymb,amsthm,mathrsfs,amsfonts,dsfont} 
\DeclareMathAlphabet{\mathpzc}{OT1}{pzc}{m}{it}
\usepackage{algorithm,algorithmic,setspace}
\usepackage{graphicx}
\usepackage{textcomp}
\usepackage{xcolor}
\usepackage{hyperref}
\usepackage{tabularx,booktabs}
\usepackage{adjustbox}
\usepackage{array,etoolbox}
\usepackage{dcolumn}
\usepackage{rotating}
\usepackage{pifont}
\usepackage{tikz}
\usepackage{comment}
\usepackage{dsfont}
\usepackage{multicol}
\usepackage{multirow}
\usepackage{mathtools}
\usepackage{stfloats}
\usepackage{lettrine}

\usepackage[nopostdot,style=super,nonumberlist, toc]{glossaries}
\usepackage{tabularx,booktabs}
\usepackage[utf8]{inputenc}
\usepackage{fourier} 
\usepackage{array}
\usepackage{makecell}
\newcolumntype{C}{>{\centering\arraybackslash}X} 

\title{QoS-Aware State-Augmented Learnable Framework for 5G NR-U/Wi-Fi Coexistence:
    Impact of Parameter Selection and Enhanced Collision Resolution
\thanks{This work has been accepted for publication in ICNC 2026. Copyright may be transferred without notice, after which this version may no longer be accessible.}
}

\author{
  Mohammad Reza Fasihi, Brian L. Mark \\
  Dept. of Electrical and Computer Engineering and Wireless Cyber Center \\
  George Mason University \\
  Fairfax, Virginia, United States\\
  \texttt{\{mfasihi4, bmark\}@gmu.edu} \\
}

\date{15 December 2025}

\begin{document}
\maketitle

\begin{abstract}
Unlicensed spectrum supports diverse traffic with stringent Quality-of-Service (QoS) requirements. In NR-U/Wi-Fi coexistence, 
the values of MAC parameters critically influence delay, collision behavior, and airtime fairness and efficiency. 
In this paper, we investigate the impact of (i) cost scaling and violation modeling, (ii) choice of MAC
parameters, and (iii) an enhanced collision resolution scheme for the Listen-Before-Talk (LBT) mechanism on the performance
of a state-augmented constrained reinforcement learning controller for NR-U/Wi-Fi coexistence. Coexistence control is formulated as a constrained Markov decision process with an explicit delay constraint for high-priority traffic and fairness as the optimization goal. Our simulation results show three key findings: (1) signed, threshold-invariant cost scaling with temporal smoothing stabilizes learning and strengthens long-term constraint adherence; (2) use of the contention window parameter for control provides smoother adaptation and better delay compliance than other MAC parameters; and (3) enhanced LBT significantly reduces collisions and improves airtime efficiency. These findings provide practical insights for achieving robust, QoS-aware coexistence control.
\end{abstract}

\keywords{
5G NR-U \and Wi-Fi \and coexistence \and unlicensed spectrum \and reinforcement learning \and quality of service (QoS)
}

\section{Introduction}
\label{sec:introduction}

Unlicensed spectrum plays a key role in supporting diverse wireless services. With 5G New Radio in Unlicensed spectrum (NR-U) extending operation into frequency bands already used by Wi-Fi, coexistence management must ensure strict Quality-of-Service (QoS) for high-priority traffic while maintaining fair and efficient spectrum sharing. This challenge arises because medium access is decentralized and highly sensitive to MAC-layer parameters since small changes in these parameters can substantially alter collision probability, access delay, and airtime fairness across coexisting technologies and traffic classes~\cite{Sathya:2020, Saha:2021, Hirzallah:2021}.

Reinforcement learning (RL) has emerged as a promising approach for adaptive coexistence optimization. 
In \cite{Wang:2021}, an RL approach is introduced for joint allocation of transmission opportunities and spectrum in NR-U/Wi-Fi coexistence to achieve efficient and fair spectrum sharing. 
An analytical study of LTE-LAA and Wi-Fi coexistence with multiple priority classes is presented in \cite{Hirzallah:2019}. In~\cite{Zou:2023}, coexistence in the presence of URLLC traffic is
addressed using a mixed-priority scheduling mechanism that balances latency and data-size demands. 
The state-augmentation method for constrained~RL is proposed in \cite{Fullana:2021} to address the limitations of traditional primal–dual and regularized approaches. The same idea is applied in \cite{NaderiAlizadeh:2022} for constrained resource management in multi-user wireless networks. 

Building on our earlier work on a QoS-aware State-Augmented Learnable (QaSAL) framework for coexistence management~\cite{Fasihi:2024,Fasihi:2025,Fasihi_2:2025}, the present paper focuses on the impact of
several enhancements. The main contributions are threefold. First, we study the impact of cost scaling and violation modeling in constrained RL, demonstrating how proper signal normalization and smoothing can stabilize learning and enhance long-term constraint compliance. Second, we investigate the effect of controlling different coexistence parameters on QoS adherence and show how their selection influences the agent’s ability to meet delay constraints for high-priority traffic. Third, we analyze the impact of enhanced contention mechanisms, in particular,
the collision resolution LBT (CR-LBT) scheme proposed by~\cite{Loginov:2021}, on spectrum efficiency and collision probability under dynamic traffic loads. These findings provide practical insights into the design of
RL-based coexistence controllers capable of ensuring reliable performance in unlicensed spectrum environments.

The remainder of the paper is organized as follows. Section~\ref{sec:system_model} provides an overview of the considered system
model. Section~\ref{sec:qasal_framework} outlines the QaSAL framework and the enhancements studied
in this paper. Section~\ref{sec:simulation_results} presents simulations results. Conclusions are given in Section~\ref{sec:conclusion}.
\section{System Model}
\label{sec:system_model}

We consider a 5G NR-U network coexisting with Wi-Fi on a shared 5~GHz unlicensed channel, where transmitters from both systems contend for downlink access to the same medium. Unless stated otherwise, traffic is saturated to represent a stressed coexistence scenario.
The Wi-Fi network follows IEEE 802.11 EDCA with binary exponential backoff, in which four access categories use distinct contention parameters~\cite{ETSI_TS_137213}. Similarly, NR-U employs Listen-Before-Talk (LBT) with four priority classes aligned with Wi-Fi’s categories. Since NR-U operates with slot-based scheduling, transmissions can begin only at slot boundaries. When the channel is sensed idle after backoff, the gNB transmits a reservation signal (RS) until the next slot to prevent Wi-Fi access. 
Each network supports two traffic priorities: PC1 (high priority, strict delay) and PC3 (low priority, relaxed delay). The stochastic nature of contention-based access can degrade PC1 latency, particularly under cross-technology interference, and delay variability is more pronounced than in licensed NR~\cite{Le:2021}.

In~\cite{Fasihi_2:2025}, the QaSAL controller dynamically adjusted the CW (Contention Window) parameter, which defines
the backoff range. In this paper, we consider two additional key MAC-layer coexistence parameters: (i) AIFSN (Arbitration Inter-Frame Spacing Number), which determines the deferment period before backoff; and (ii) MCOT (Maximum Channel Occupancy Time), which limits the maximum channel occupancy time. Adjusting any of these three parameters directly impacts collision rate, access delay, airtime efficiency, and fairness. In addition, the impact of an enhanced contention mechanism for 5G~NR-U proposed
by~\cite{Loginov:2021}, called collision resolution LBT (CR-LBT), is studied. CR-LBT turns the brief pre-transmission gap into a few quick pulse-and-listen checks, letting the gNB avoid stepping on nodes still counting down and cut simultaneous-start collisions, improving efficiency under heavy load.

The QaSAL framework is implemented in a SimPy-based discrete-event  simulator that models sensing, backoff, collisions, reservation time, and successful transmissions. Two performance indicators are evaluated: the medium-access delay of PC1 traffic, representing the QoS constraint, and airtime fairness measured by Jain’s Fairness Index. Coexistence Parameter Management (CPM) refers to the adaptive tuning of MAC parameters to meet QoS targets for high-priority traffic while maintaining fair spectrum sharing. 

\section{QaSAL Framework for CPM and Enhancements}
\label{sec:qasal_framework}

\subsection{CPM as Constrained MDP}

The CPM problem is formulated as a constrained Markov decision process (CMDP). The goal is to learn a control policy that adjusts coexistence parameters 
to satisfy QoS constraints for high-priority traffic while maximizing overall airtime fairness among nodes from different priority classes.
Let $\mathpzc{S}\subset\mathds{R}^n$ denote a compact set of coexistence states describing the joint NR-U/Wi-Fi environment. At time $t$, given $\mathbf{S}_t\in\mathpzc{S}$, the agent chooses a CPM action $\mathbf{a}_t=\mathbf{a}(\mathbf{S}_t)\in\mathds{R}^a$ under policy $\pi$. The system evolves with transition probability $p(\mathbf{S}_{t+1}|\mathbf{S}_t,\mathbf{a}_t)$ and returns a vector of performance signals $\mathbf{f}(\mathbf{S}_t,\mathbf{a}_t)\in\mathds{R}^m$ (e.g., PC1 delay, collision rate, airtime fairness). Long-term performance is evaluated via \textit{ergodic averages} defined as follows:
\begin{equation}
    \Tilde{V}_i(\pi)=\frac{1}{T}\sum_{t=0}^{T-1} f_i(\mathbf{S}_t,\mathbf{a}_t),\quad i=0,\dots,m-1.
    \label{eq:ErgodicValueFunction}
\end{equation}

Since NR-U and Wi-Fi operate under different access mechanisms and traffic priorities, their objectives often conflict. For example, minimizing delay for high-priority traffic may reduce overall fairness or throughput. Therefore, CPM is a multi-objective optimization problem which can be solved through multi-objective optimization approaches. A common approach is to combine multiple objectives into a single objective using the \textit{linear scalarization} approach. Although this approach is simple and computationally efficient, it relies on predefined weights that may not accurately capture nonlinear trade-offs among objectives. 

The QaSAL approach adopts a constrained RL formulation. In this setting, the first objective $f_0(\mathbf{S}_t,\mathbf{a}(\mathbf{S}_t))$, is treated as the \textit{primary} objective to maximize, while the remaining objectives serve as \textit{QoS constraints}. The goal is to find the CPM decision vector $\mathbf{a}(\mathbf{S}_t)$ that optimizes the primary objective while satisfying the constraints for every environment state $\mathbf{S}_t\in\mathpzc{S}$. The corresponding parameterized constrained CPM problem is formulated as
\begin{align}
    \max_{\boldsymbol{\theta}\in\boldsymbol{\Theta}} \quad &
        \frac{1}{T}\sum_{t=0}^{T-1} f_0(\mathbf{S}_t, \mathbf{a}(\mathbf{S}_t;\boldsymbol{\theta})),
        \label{eq:ParameterizedCPM:Objective} \\
    \text{s.t.} \quad &
        \frac{1}{T}\sum_{t=0}^{T-1} f_i(\mathbf{S}_t, \mathbf{a}(\mathbf{S}_t;\boldsymbol{\theta}))
        \geq c_i, \quad i=1,\dots,m-1,
        \label{eq:ParameterizedCPM:Constraints}
\end{align}
where $c_i\in\mathds{R}$ denotes the threshold for the $i$-th QoS metric. 

\subsection{State-Augmented RL for CPM  and QaSAL}

The \textit{state-augmentation approach}~\cite{Fullana:2024} provides a practical alternative to traditional primal–dual approach for solving constrained RL problems by embedding the dual variables directly into the agent’s state. This converts the problem into a state-augmented MDP, where constraint satisfaction becomes part of the environment’s dynamics. As the dual variables evolve online, the policy remains constraint-aware and adapts to varying traffic and contention conditions without repeated re-optimization. 

Let $\mathbf{S}_t$ denote the environment state at time step $t$ in the $k$-th epoch, and $\boldsymbol{\lambda}_k$ the vector of dual variables. Augmenting $\boldsymbol{\lambda}_k$ into the state yields the augmented state 
$\Tilde{\mathbf{S}}_t=(\mathbf{S}_t,\boldsymbol{\lambda}_k)$. 
The CPM policy is then parameterized as $\mathbf{a}(\Tilde{\mathbf{S}}_t;\boldsymbol{\Tilde{\theta}})$, where $\boldsymbol{\Tilde{\theta}}\in\boldsymbol{\Tilde{\Theta}}$ denotes the parameter of the state-augmented policy. 
The augmented Lagrangian is defined as
\begin{align}
    \mathcal{L}(\boldsymbol{\lambda};\boldsymbol{\Tilde{\theta}}) 
    &= \frac{1}{T}\sum_{t=0}^{T-1} f_0(\Tilde{\mathbf{S}}_t, \mathbf{a}(\Tilde{\mathbf{S}}_t;\boldsymbol{\Tilde{\theta}}))  \nonumber \\
    &\quad + \sum_{i=1}^{m-1}\lambda_i
    \!\left(\frac{1}{T}\sum_{t=0}^{T-1} f_i(\Tilde{\mathbf{S}}_t, \mathbf{a}(\Tilde{\mathbf{S}}_t;\boldsymbol{\Tilde{\theta}})) - c_i\right).
    \label{eq:ParameterizedAugmentedLagrangian}
\end{align}

Given a probability distribution $p_{\boldsymbol{\lambda}}$ for the dual variables, the optimal state-augmented CPM policy maximizes the expected augmented Lagrangian:
\begin{equation}
    \boldsymbol{\Tilde{\theta}}^* =
    \arg\max_{\boldsymbol{\Tilde{\theta}}\in\boldsymbol{\Tilde{\Theta}}}
    ~\mathds{E}_{\boldsymbol{\lambda}\sim p_{\boldsymbol{\lambda}}}
    \!\left\{\mathcal{L}(\boldsymbol{\lambda};\boldsymbol{\Tilde{\theta}})\right\}.
    \label{eq:MinMaxAugmentedProblem}
\end{equation}
Using the optimal parameter $\boldsymbol{\Tilde{\theta}}^*$, the CPM decisions 
$\mathbf{a}(\Tilde{\mathbf{S}};\boldsymbol{\Tilde{\theta}})$ maximize the Lagrangian for the current $\boldsymbol{\lambda}$ at each iteration $k$. 
The dual variables are then updated by
\begin{equation}
    \lambda_{i,k+1} =
    \left[\lambda_{i,k}-
    \frac{\eta_{\lambda_i}}{T_0}
    \sum_{t=kT_0}^{(k+1)T_0-1}
    \!\left(f_i(\Tilde{\mathbf{S}}_t, \mathbf{a}(\Tilde{\mathbf{S}}_t;\boldsymbol{\Tilde{\theta}}_k^*)) - c_i\right)
    \right]^+,
    \label{eq:AugmentedLambdaUpdateIteration}
\end{equation}
for $i=1,\dots,m-1$. In CPM, where traffic priorities and contention vary rapidly, state-augmentation keeps the policy aligned with real-time constraints, improving compliance for high-priority traffic while maintaining fair  coexistence. The QaSAL framework implementing this concept was introduced in our earlier work~\cite{Fasihi_2:2025} where we applied both the primal-dual and QaSAL approaches to the coexistence of NR-U/Wi-Fi and compared there performance.

\begin{figure*}[t]
\centering
    \includegraphics[width=0.95\linewidth]{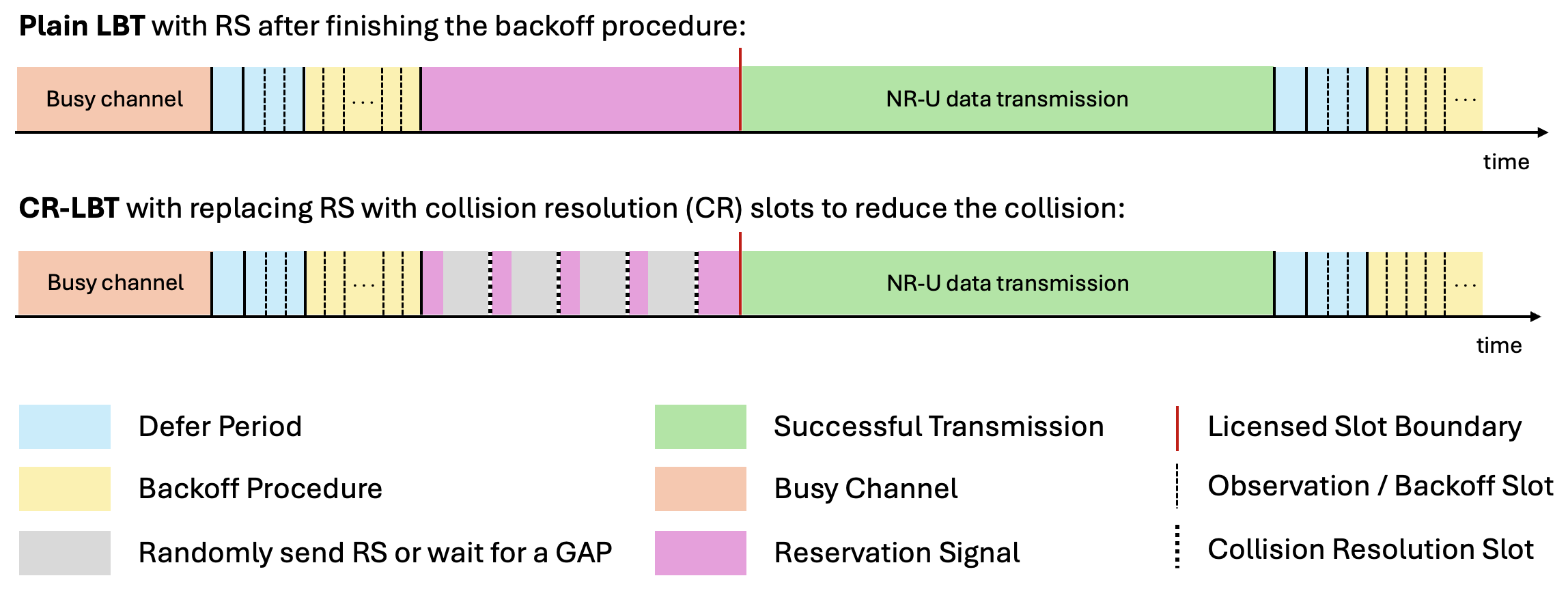}
    \caption{Enhanced LBT in NR-U gNBs. Plain LBT transmits a reservation signal (RS) after backoff until the next slot, whereas 
    CR-LBT~\cite{Loginov:2021} replaces RS with short collision-resolution (CR) slots that reduce collisions while keeping slot alignment.}
    \label{fig:cr_lbt}
\end{figure*}

For the NR-U/Wi-Fi coexistence scenario, the network state $\mathbf{S}_t$ includes the PC1 medium-access delay (instantaneous and smoothed), collision rates and short-term collision trends, channel airtime utilization, and QoS-violation rate. These features provide the agent with both current network conditions and recent contention dynamics relevant to coexistence.
At each step, the CPM action is $\mathbf{a}_t=\mathbf{a}(\mathbf{S}_t)$, which controls a particular MAC parameter. As discussed in Section~\ref{sec:system_model}, we consider the impact
of using the parameters CW, AIFSN, and MCOT.
The constrained CPM problem is then given by
\begin{equation}
\max_{\{\mathbf{a}_t\}_{t=0}^{T-1}} \frac{1}{T}\sum_{t=0}^{T-1}\mathrm{JFI}_t 
\quad \text{s.t.} \quad \frac{1}{T}\sum_{t=0}^{T-1}\overline{D}_{\text{PC1},t} \le D_{\text{th}}.
\label{eq:ParameterizedCPMCoexistence}
\end{equation}
The dual variable associated with the delay constraint is updated online and appended to the state, enabling the policy to adapt in real time to traffic and contention dynamics while maintaining the PC1 delay requirement. 

\subsection{Cost-Scaling and Violation Modeling}

In constrained RL, the main challenge is turning a noisy and time-varying constraint signal into a stable learning target while optimizing the primary objective. This challenge is particularly acute in 5G NR-U/Wi-Fi coexistence, where decentralized access, collisions, and reservation signaling cause bursty delays and a small feasible region. Even small errors in violation feedback can destabilize learning. By embedding the dual variable in the state, state-augmentation makes the constraints visible to the policy. 


\subsubsection{Signed, Threshold-Invariant Scaling}

Learning stability depends strongly on how the constraint signal is represented. We employ a signed, threshold-invariant scaling based on a smooth hyperbolic tangent function. Let
\[
v \triangleq \frac{f_i(\Tilde{\mathbf{S}}_t, \mathbf{a}(\Tilde{\mathbf{S}}_t;\boldsymbol{\Tilde{\theta}})) - c_i}{c_i},
\]
denote the signed relative violation, where \(v > 0\) means the constraint is satisfied. To bound the signal and preserve smooth gradients, we apply a scaled \(\tanh\) transformation:
\[
v_{\text{scaled}} = \tanh\!\left(\frac{v}{\kappa}\right),
\]
where \(\kappa\) controls the transition slope near the threshold. Smaller \(\kappa\) increases sensitivity to violations, while larger values yield smoother but less responsive updates. The agent observes only the negative component \(v^{-} = \min(0,\, v_{\text{scaled}})\) so that it penalizes falling below the constraint but not safe operation above it. Meanwhile, the dual variable and the state incorporate the full signed \(v_{\text{scaled}}\), enabling the policy to tighten or relax constraint pressure dynamically and maintain stable learning behavior.

\subsubsection{Consistent dual update}

To prevent scale mismatch between training and deployment, the dual update uses the same scaled and smoothed signal as the learner. Every $T_0$ steps,
\[
\lambda \leftarrow \Big[\lambda - \eta_{\lambda}\,\overline{v_{\text{clip}}}\Big]^+,\qquad \lambda\in[0,\lambda_{\max}],
\]
where $\overline{v_{\text{clip}}}$ is an exponential moving average of $v_{\text{clip}}$. Matching the smoothing and normalization across learning and dual updates avoids inconsistencies that cause oscillations. The update rate $(\eta_{\lambda},T_0)$ should reflect contention dynamics; heavier loads may require larger $\eta_{\lambda}$ and smaller $T_0$. If $\lambda$ saturates at $\lambda_{\max}$ without reducing violations, $\lambda_{\max}$ can be increased or $\kappa$ can be slightly decreased.

Overall, the signed, threshold-invariant scaling provides smooth, well-conditioned gradients around the constraint boundary. Negative-only costs keep the policy efficient, while signed signals allow the dual to adapt pressure dynamically. Using the same scaled and smoothed signal in both learning and dual updates improves long-term constraint adherence without complicated reward shaping.

\subsection{Collision Resolution for 5G NR-U/Wi-Fi Coexistence}

Collisions in unlicensed bands arise from heterogeneous carrier-sense thresholds, asynchronous timing, and partial visibility between NR-U and Wi-Fi devices. Misaligned slot boundaries and long reservation intervals can cause overlapping transmissions and extra delay, especially for high-priority traffic, while also compromising overall fairness. 
Enhanced Listen-Before-Talk (LBT) mechanisms address these issues by turning short pre-transmission gaps into decision points for collision avoidance. In particular, collision-Resolution LBT (CR-LBT) interleaves brief sensing opportunities before slot boundaries, allowing gNBs to detect active transmissions and defer access when necessary. As illustrated in Fig.~\ref{fig:cr_lbt}, CR-LBT replaces the reservation signal with short collision-resolution slots, reducing collisions and improving airtime efficiency under heavy contention. For further details on CR-LBT, refer to~\cite{Loginov:2021}.

\subsection{Computational Complexity at Deployment}

%
During the online execution phase, QaSAL uses a fixed policy obtained in the offline training phase and performs only a forward pass through the Q-network followed by an argmax operation over the discrete action set. The per-step cost is therefore a single inference of $O(P)$ multiply-accumulate operations for a network with $P$ parameters, with no optimizer state, no replay buffer, and predictable latency. In contrast, the primal-dual method continues to adapt in the execution phase by updating the dual variables $\lambda$ as well as the value function to stay aligned with the changing augmented objective $r - \lambda v$. This involves gradient evaluations, target computations, and memory traffic, resulting in much higher 
and more variable per-step computational cost compared to QaSAL. 

\begin{table}[t]
    \centering
    \caption{Simulation Setup}
    \renewcommand{\arraystretch}{1}
    {\small
    \begin{tabular}{ c  c }
        \Xhline{2\arrayrulewidth}
        Parameter & Value \\ 
        \hline
        Interaction time & $10,000$~episodes \\
        Episode duration & $100$~steps \\
        Step duration & $2.5~\text{ms}$ \\
        Discount factor & $0.99$ \\ 
        Replay buffer size & 100,000 \\
        Range of $\epsilon$ & 1.0 to 0.01 \\ 
        DQN learning rate & $10^{-5}$ \\
        Batch size & 256 \\
        Hidden layers & $3 \times 1024$\\
        $\lambda_{\text{max}}, T_0, \eta_{\lambda}, \kappa$ & 5.0, 5, 0.05, 0.5\\
        $D_{\text{th}}$ & $2~\text{ms}$\\
        \Xhline{2\arrayrulewidth}
    \end{tabular}
    }
    \label{tab:DQN_param}
\end{table}

\section{Simulation Results}
\label{sec:simulation_results}

In our previous work~\cite{Fasihi:2025, Fasihi_2:2025}, we showed that QaSAL has 
clear advantages over classical primal–dual and linearly scalarized RL approaches by achieving smoother training dynamics and stronger constraint satisfaction in NR-U/Wi-Fi coexistence. The present study focuses 
on the behavior of QaSAL under different design and control choices. The step duration is selected 
to be large enough to include several transmission attempts to enable accurate calculation of the medium access delay. 

\subsection{Experimental Setup}

The hyperparameters used for the simulation study are summarized in Table~\ref{tab:DQN_param}.
At each step, the CPM action $\mathbf{a}_t=\mathbf{a}(\mathbf{S}_t)$ depends on the MAC parameter that is controlled. When controlling the \textit{contention window (CW)}, $\mathbf{a}_t\in\{0,1,\ldots,6\}$ and the maximum contention window for priority class $i\!\in\!\{\text{PC1},\text{PC3}\}$ is
\[
\mathrm{CW}_{\max,\text{PC}i}=2^{\,a_{i,t}+b_i}-1,
\quad \text{with } b_{\text{PC1}}=0,\; b_{\text{PC3}}=4,
\]
which directly affects backoff duration and channel access aggressiveness. 
For \textit{AIFSN} control, the action selects from predefined sets 
$\text{AIFS}_{\text{PC1}}\!\in\![1,2,3]$ and $\text{AIFS}_{\text{PC3}}\!\in\![1,\ldots,7]$, 
tuning the deferment period before backoff and thus the relative priority among transmitters. 
When \textit{MCOT} is the control parameter, the agent selects 
$\text{MCOT}_{\text{PC}i}\!\in\![1000,1500,\ldots,4000]~\mu\text{s}$, 
which governs the maximum duration a transmitter may occupy the channel after winning access. 
These actions collectively determine how aggressively each node contends for the medium and how long it holds the channel, jointly shaping delay, collision probability, and fairness. We consider
\[
f_0(\mathbf{S}_t,\mathbf{a}(\mathbf{S}_t))=\mathrm{JFI}_t, 
\qquad
f_1(\mathbf{S}_t,\mathbf{a}(\mathbf{S}_t))=\overline{D}_{\text{PC1},t},
\]
where $\mathrm{JFI}_t$ measures airtime fairness (PC1/PC3) and $\overline{D}_{\text{PC1},t}$ is the smoothed medium-access delay for PC1.

\subsection{Effect of Violation Modeling and Cost Scaling}
\label{subsec:violation_results}

\begin{figure*}
\begin{multicols}{2}
\centering
    \includegraphics[width=0.96\linewidth]{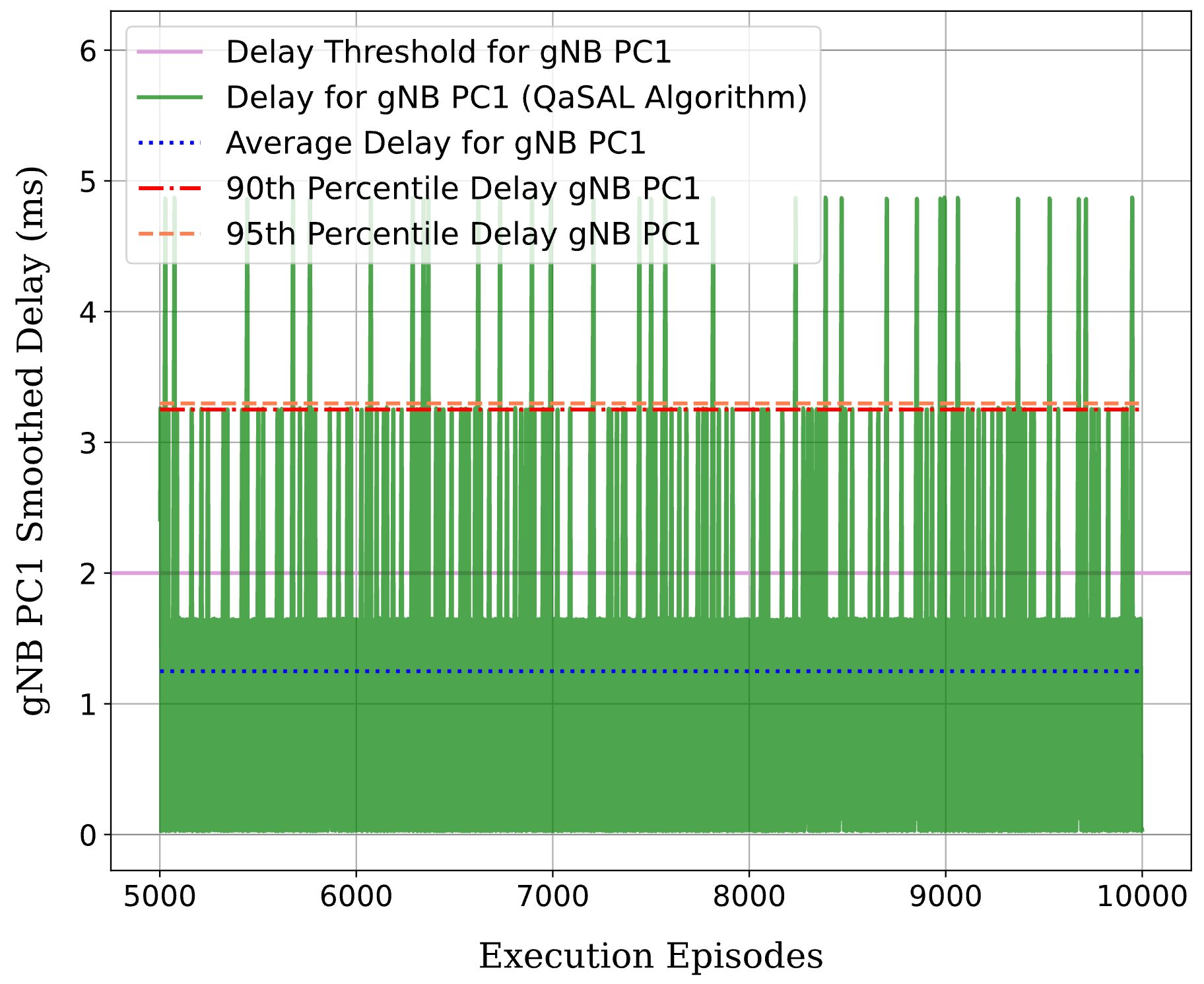}
    \caption{QaSAL without violation modeling and cost scaling; action is CW.}
    \label{fig:QaSAL_no_scaling}
    \par
	\includegraphics[width=0.96\linewidth]{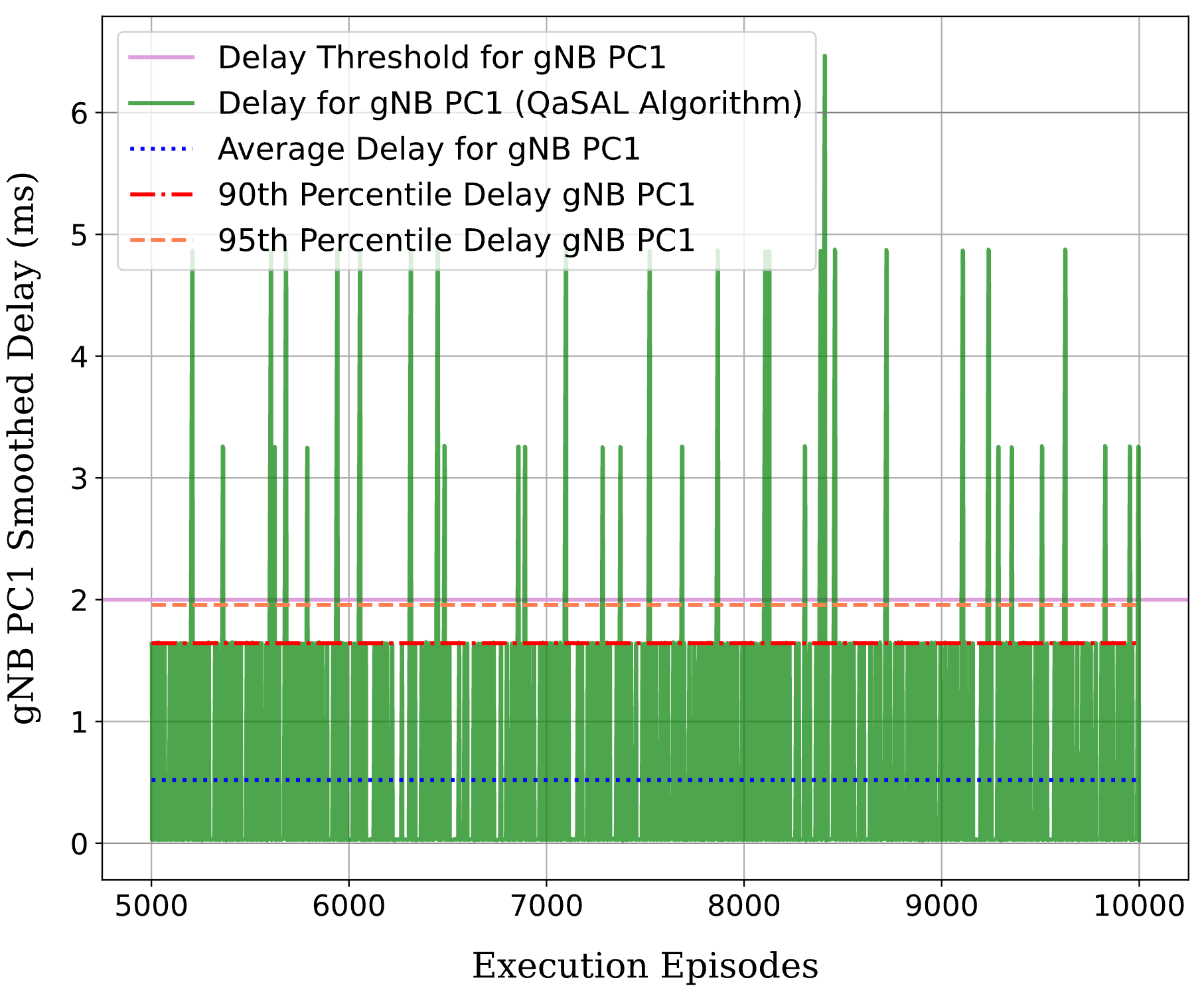}
    \caption{QaSAL with violation modeling and cost scaling; action is CW.}
    \label{fig:QaSAL_with_scaling}
    \par
\end{multicols}
\vspace*{-0.15in}
\end{figure*}

Figs.~\ref{fig:QaSAL_no_scaling} and~\ref{fig:QaSAL_with_scaling}, respectively, compare the delay performance of the QaSAL controller without and with cost scaling and violation modeling when the action controls CW. Without scaling, the delay signal shows frequent spikes above the delay threshold, and the 95th-percentile delay remains above the threshold. This behavior indicates that the dual update reacts too sharply to raw violation signals, causing oscillations and weak adherence to the QoS constraint.
When cost scaling and violation modeling are applied, the smoothed delay stays consistently below the threshold with fewer and shorter violations, showing that the agent learns to balance fairness and constraint satisfaction more effectively. Signed, threshold-invariant scaling combined with temporal smoothing produces a well-conditioned feedback signal. This improves convergence stability and long-term compliance while preserving delay performance.

\begin{figure*}
\begin{multicols}{2}
\centering
	\includegraphics[width=0.96\linewidth]{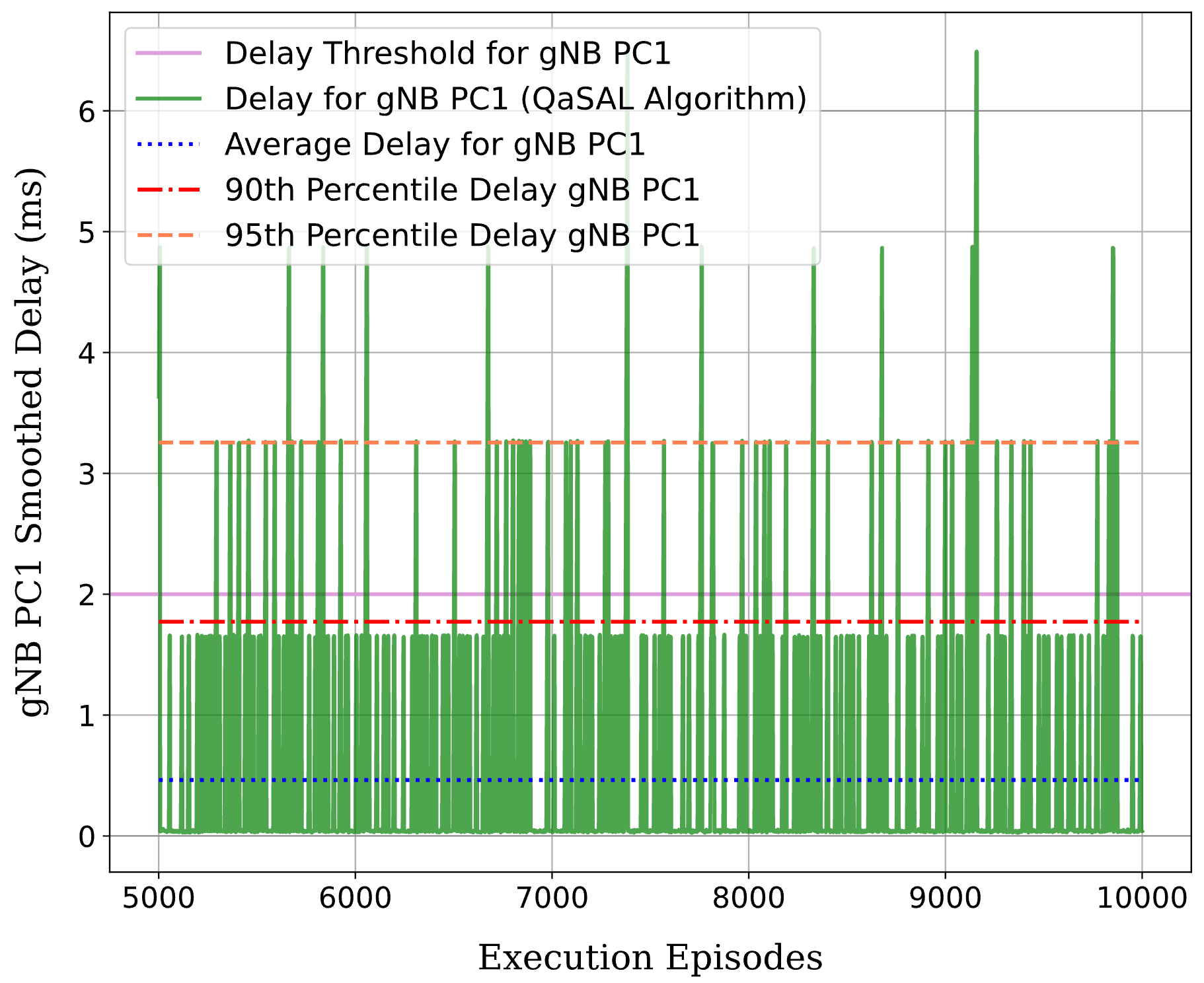}
    \caption{QaSAL with violation modeling and cost scaling; action is AIFSN.}
    \label{fig:QaSAL_with_AIFSN}
    \par
    \includegraphics[width=0.96\linewidth]{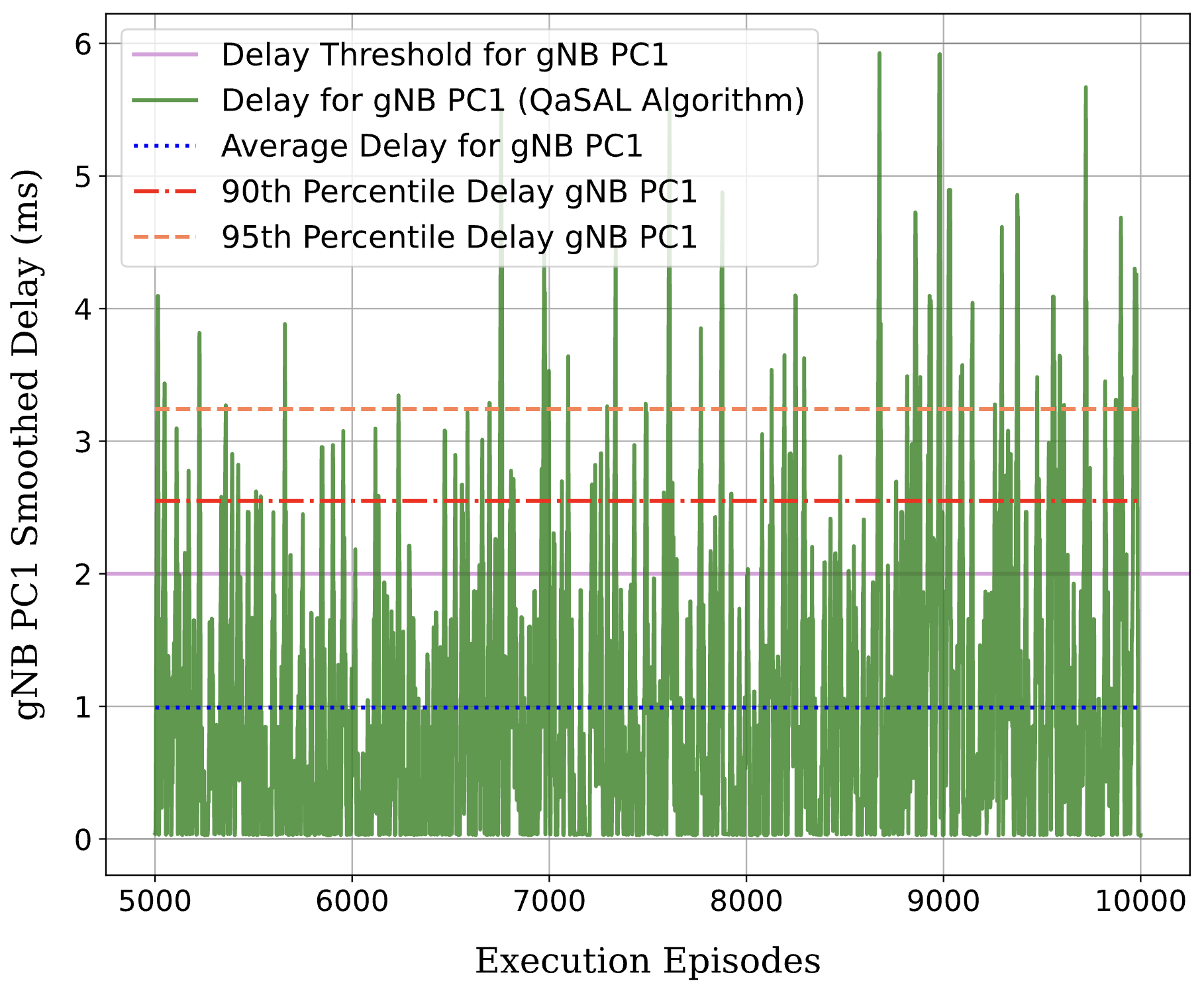}
    \caption{QaSAL with violation modeling and cost scaling; action is MCOT.}
    \label{fig:QaSAL_with_MCOT}
    \par
\end{multicols}
\vspace*{-0.15in}
\end{figure*}

\subsection{Impact of Coexistence Parameter Selection}
\label{subsec:parameter_results}

Figure~\ref{fig:QaSAL_with_AIFSN} shows QaSAL performance when the action controls AIFSN. Compared with the CW-controlled case, AIFSN yields more frequent delay spikes, and the 95th-percentile delay sits above the threshold. Two issues likely cause this result: (i) AIFSN changes the fixed deferment period before backoff begins, causing class-wide timing alignment shifts that can suddenly increase contention in the early slots; and (ii) the available AIFSN levels are coarse, so single-step policy changes cause large priority shifts and not just statistical spacing. In contrast, CW offers a finer ``gain'' on contention aggressiveness, smoothing access without amplifying burstiness. Therefore, under our traffic load and priority mix, CW is the more effective primary knob for delay constraint adherence; AIFSN should be used cautiously to avoid tail inflation. Fig.~\ref{fig:QaSAL_with_MCOT} shows QaSAL with MCOT as the action. Since MCOT changes how long the transmitter holds the channel after already winning the contention, it is apparently
a weaker lever on medium access delay, especially under saturation conditions. 

\subsection{Effect of Enhanced LBT Mechanism (CR-LBT)}
\label{subsec:crlbt_results}

\begin{figure}[t]
    \centering
    \includegraphics[width=0.55\linewidth]{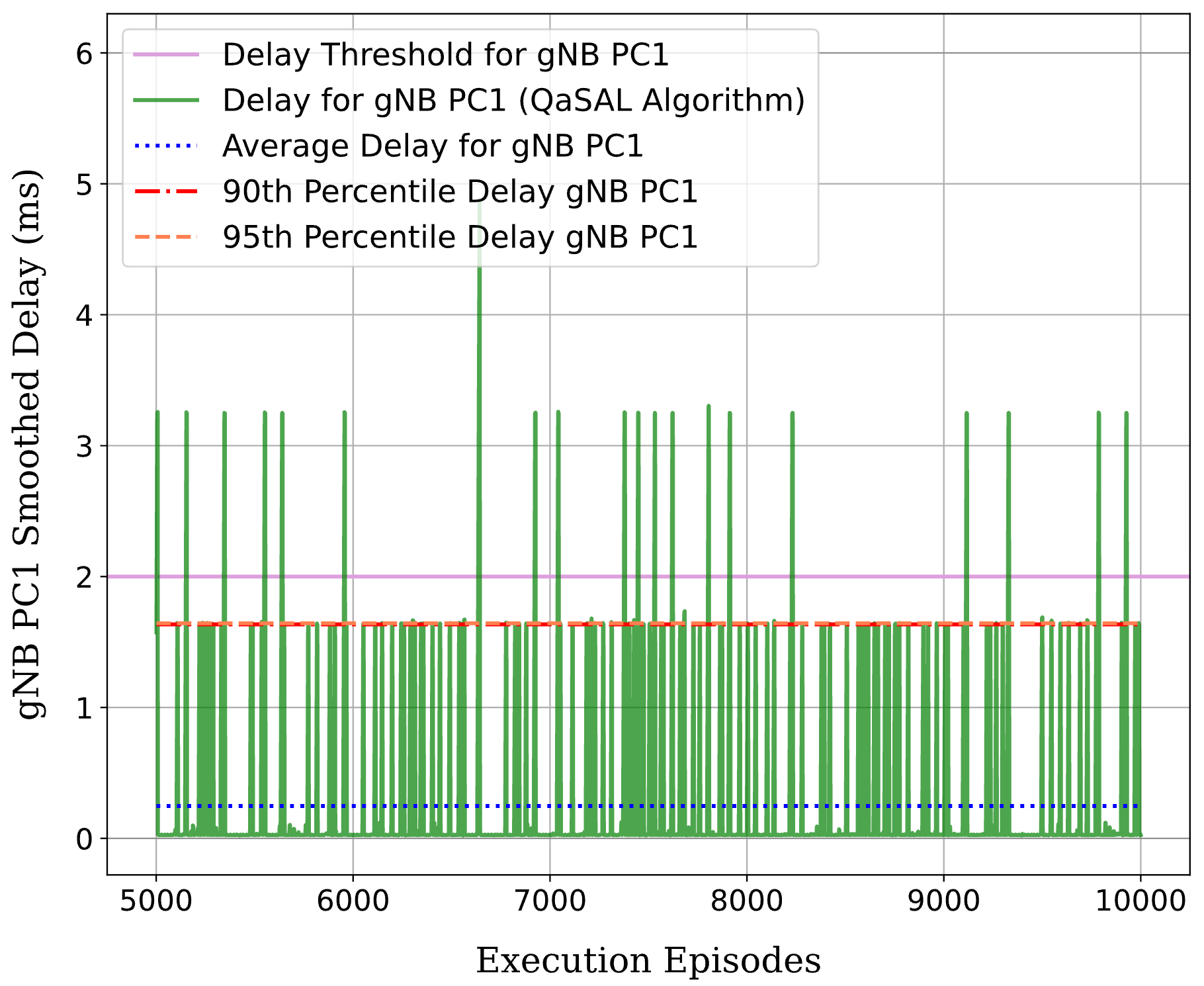}
    \caption{QaSAL with violation modeling, cost scaling, and CR-LBT; action is CW.}
    \label{fig:QaSAL_with_crlbt}
\end{figure}

Fig.~\ref{fig:QaSAL_with_crlbt} illustrates the delay performance of the QaSAL controller when CR-LBT is enabled alongside cost scaling and violation modeling. Compared with plain LBT (Fig.~\ref{fig:QaSAL_with_scaling}), CR-LBT significantly reduces both the frequency and magnitude of delay spikes, keeping the average and 95th-percentile delays consistently below the threshold. This improvement stems from the additional collision-resolution slots, which mitigate cross-technology collisions and contention near slot boundaries.
Moreover, CR-LBT enhances channel utilization for all nodes—particularly those carrying low-priority traffic—by reducing the overall collision probability. Table~\ref{tab:qasal_crlbt_comparison} summarizes the performance across different network components under QaSAL with plain LBT and CR-LBT. The results show that enhanced collision resolution effectively complements QoS-aware learning by lowering retransmissions and improving overall airtime efficiency.

\begin{table}[t]
\centering
\caption{QaSAL performance with Plain LBT vs.\ CR-LBT.}
\label{tab:qasal_crlbt_comparison}
\renewcommand{\arraystretch}{0.93}
\small
\begin{tabular}{lccc}
    \toprule
    \multicolumn{4}{c}{Average Collision Probability}\\
    \midrule
    Node & Plain LBT & CR-LBT & $\Delta$\\
    \midrule
    gNB PC1 & 10.7\% & 0.25\%  & $-10.45\%$ \\
    gNB PC3 & 97.4\% & 45.7\% & $-51.7\%$ \\
    AP  PC3 & 98.6\% & 61.6\%  & $-37.0\%$ \\
    \midrule
    \multicolumn{4}{c}{Average Airtime Efficiency}\\
    \midrule
    Node & Plain LBT & CR-LBT & $\Delta$\\
    \midrule
    gNB PC1 & 89.0\% & 100.0\% & $+11.0\%$ \\
    gNB PC3 & 9.0\% & 56.0\% & $+47.0\%$ \\
    AP  PC3 & 8.0\% & 32.0\% & $+24.0\%$ \\
    \midrule
    \multicolumn{4}{c}{Average Delay}\\
    \midrule
    Node & Plain LBT & CR-LBT & $\Delta$\\
    \midrule
    gNB PC1 & 0.52~ms & 0.25~ms & $-0.27$~ms \\
    \bottomrule
\end{tabular}
\end{table}
\section{Conclusion}
\label{sec:conclusion}

We investigated the performance of QoS-aware state-augmented
constrained RL for NR-U/Wi-Fi coexistence~\cite{Fasihi:2025} with several
enhancements and different MAC-layer parameter choices. 
Our simulation results showed that cost scaling and violation modeling greatly improve QaSAL’s learning stability and constraint adherence. Without scaling, the delay fluctuates and often exceeds the threshold, while the scaled version maintains smoother behavior and tighter control near the QoS threshold. Using signed violations for the dual variable and positive-only ones for the learner keeps the agent efficient and stable over long runs. Among coexistence parameters, tuning the CW parameter provides the most reliable delay control, whereas parameter control
using AIFSN and MCOT shows higher variability. The enhanced contention mechanism, CR-LBT, lowers collisions and improves airtime efficiency but has limited effect on delay compliance. The CW-based actions with CR-LBT and proper cost scaling yielded the best stability and QoS performance. Future work will extend QaSAL to multi-channel coexistence and control of PHY-layer parameters.

\bibliographystyle{IEEEtran}
\bibliography{IEEEabrv,main}
\end{document}